\documentstyle[epsfig,prl,twocolumn,aps]{revtex}

\def\Ai{\hbox{\hbox{${\cal A}$}}\kern-1.9mm{\hbox{${/}$}}}
\def\Vi{\hbox{\hbox{${\cal V}$}}\kern-1.9mm{\hbox{${/}$}}}
\def\Di{\hbox{\hbox{${\cal D}$}}\kern-1.9mm{\hbox{${/}$}}}
\def\lam{\hbox{\hbox{${\lambda}$}}\kern-1.6mm{\hbox{${/}$}}}
\def\D{\hbox{\hbox{${D}$}}\kern-1.9mm{\hbox{${/}$}}}
\def\A{\hbox{\hbox{${A}$}}\kern-1.8mm{\hbox{${/}$}}}
\def\V{\hbox{\hbox{${V}$}}\kern-1.9mm{\hbox{${/}$}}}
\def\parz{\hbox{\hbox{${\partial}$}}\kern-1.7mm{\hbox{${/}$}}}
\def\B{\hbox{\hbox{${B}$}}\kern-1.7mm{\hbox{${/}$}}}
\def\R{\hbox{\hbox{${R}$}}\kern-1.7mm{\hbox{${/}$}}}
\def\si{\hbox{\hbox{${\xi}$}}\kern-1.7mm{\hbox{${/}$}}}

\begin{document}
\draft

\twocolumn[\hsize\textwidth\columnwidth\hsize\csname @twocolumnfalse\endcsname

\title{Gauge field theory of transport and magnetic relaxation in 
 underdoped cuprates }
\author{P.A. Marchetti,$^a$ Jian-Hui  Dai,$^{b,c}$   Zhao-Bin Su,$^d$ Lu Yu$^{b,d}$}

\address{ $^a$ Dipartimento di Fisica ``G. Galilei",
INFN, I--35131 Padova, Italy\\
$^b$International Centre for Theoretical Physics, I-34100 Trieste, Italy\\
$^c$Zhejiang Institute of Modern Physics, Zhejiang University, Hangzhou 310027, China\\
$^d$Institute of Theoretical Physics, CAS, Beijing 100080, China}

\maketitle

\begin{abstract}
Based on  recently proposed $U(1)\times SU(2)$ Chern-Simons gauge 
field theory, an interpretation
of the transport and magnetic relaxation properties of underdoped
cuprates is proposed, taking into account the short range antiferromagnetic
order.  The interplay of the doping-dependent spin gap (explicitly
derived by us)  effect and dissipation due to gauge fluctuations
 gives rise to a crossover from metallic  to insulating 
behavior of conductivity as  temperature decreases, in semi-quantitative  agreement
with experimental data. For the same reason the magnetic relaxation rate
shows a maximum nearby. Various crossover temperatures
related to spin gap effects are shown to be different manifestations of the same energy scale.
\end{abstract}

]


\narrowtext

The understanding  of the anomalous normal state properties of oxide
superconductors has been a challenge for theorists since their
discovery.$^{\cite{andb}}$  Recently great attention has been focused on
underdoped superconductors,$^{\cite{bal,furu,zhe,ioffe}}$  where the
pseudogap  (spin gap) effects  are essential. We will concentrate 
on the doping  range where the short range antiferromagnetic
order (SRAFO) exists,$^{\cite{kas}}$  
and propose an interpretation of the transport and magnetic relaxation
properties in this region, based on the recently proposed $U(1)\times SU(2)$
gauge field theory.$^{\cite{mar1}}$ 

The linear temperature dependence of resistivity in most of oxide superconductors
over a wide range of temperature is well established and a number of explanations have 
been proposed,$^{\cite{iye}}$ including the $U(1)$ gauge field theory.$^{\cite{ln}}$
On the other hand, in underdoped samples, a resistivity minimum and a  crossover from
metallic to insulating behavior has been observed.$^{\cite{takagi,pre,fiory}}$
A similar divergence of resistivity at low temperatures has been found
in superconducting samples in strong magnetic fields,$^{\cite{ando}}$
suppressing superconductivity. An apparently ``obvious''
explanation of these two related phenomena would be localization of 
charge carriers in two dimensions. However, a more careful comparison
of theory with  experiments shows$^{\cite{cast}}$ that the localization
effects including carrier interactions cannot correctly interpret the 
data. Several other explanations have been proposed based on
non-Fermi liquid (FL) behavior of charge carriers,$^{\cite{and,varma,mir}}$
 but the zero field experiments$^{\cite{takagi,pre,fiory}}$
have not been addressed, 
except for  \cite{mir} where a gauge field approach has been used.
We, instead, will concentrate on the latter case. We will show that the
presence of SRAFO, leading to a finite mass of spinons (bosons)
is the correct starting point in this doping range.  The self-generated
$U(1)$ holon-spinon ($h/s$) gauge field becomes singular due 
to coupling with holons (fermions),$^{\cite{ln}}$ which, in turn, renormalizes the massive 
spinons in a nontrivial way. At low temperatures,   effects due to finite
spinon mass prevail leading to insulating behavior, while at higher
temperatures the dissipation caused by the  gauge field 
 dominates and gives rise to metallic behavior.
For similar reasons, the spin relaxation rate is low at both low and high
temperatures, reaching a maximum near the resistivity crossover point
which is also consistent with   experiment.$^{\cite{ber}}$

Following a strategy previously applied to the 1D $t-J$ model which has
reproduced there the known exact Bethe Ansatz results,$^{\cite{mar1d}}$ the Chern-Simons
bosonization with $U(1)\times SU(2)$ gauge field$^{\cite{fro1}}$ 
was applied   to  the  2-D $t-J$ model  in the limit $t \gg J$, allowing 
us to rewrite 
the partition function (and the correlation functions)  in terms of a spin 
${1\over 2}$ fermion field  $\psi_\alpha, \alpha=1,2,$  minimally coupled to 
a $U(1)$  field $B$ (gauging global charge), and an $SU(2)$  field $V$
(gauging global spin) whose 
dynamics is given by a Chern--Simons (C.S.) action.$^{\cite{mar1}}$
We decomposed the fermion field $\psi_\alpha$ into  product of a spinless
fermion field $H$ (holon) and a spin $1\over 2$ boson field $\Sigma_\alpha$ 
(spinon), satisfying the constraint $\Sigma^*_\alpha \Sigma_\alpha =1$,
thus introducing a local $U(1)$ gauge invariance called $h/s$. We proved the
existence of an upper bound of the partition function for holons
in a  spinon background, and we found the  optimal spinon configuration
(s+id-like RVB state) saturating the upper bound on  average.
 After neglecting
 the feedback of holon fluctuations on  field $B$ and spinon fluctuations on   field 
$V$, the holon 
field is  a fermion and the spinon field is a hard--core boson.
Within this approximation  the ``mean field'' (MF) $\bar B$   produces a 
$\pi$ flux phase for  holons, converting them into Dirac--like fermions,
while the $\bar V$ field, taking into account the feedback of holons
  produces a gap for spinons vanishing in the 
 zero doping limit. 

The continuum action for  AF fluctuations around  the ``MF",
 described  by a spin ${1\over 2}$ boson field $z_\alpha, \alpha=1,2$
( still ``spinons'')  is given  by:$^{\cite{mar1}}$
\begin{equation}
g^{-1}\int dx_0 d^2 x [ v_s^{-2}|(\partial_0 - A_0) z |^2 
-|(\partial_\mu - A_\mu)z |^2 + m_s^2 z^*_\alpha z_\alpha], \label{spinon}
\end{equation}
where $A$ is the $h/s$ gauge field,
$g= 8 /J, \; v_s = \sqrt{2} J a$, with $a$ 
the lattice constant. 
The spinon ``mass'' term $m_s^2 \sim \langle \bar V^2 
\rangle \sim - \delta \ln \delta$ (the main novelty) 
 is due to averaged perturbation caused by holons
of concentration $\delta$ via $\bar V$. 
This explicit doping dependence was {\it derived}, rather than assumed 
in the theory. It produces a  SRAFO, with  correlation length 
$\xi_{AF} \sim (-\delta \ln \delta)^{-{1\over 2}}$,
fully consistent  with the  neutron scattering data.$^{\cite{bir}}$

Neglecting  the gauge fluctuations,
 holons are described by FL theory with a Fermi surface (FS) consisting of 4 
``half-pockets'' centered at  $(\pm {\pi\over 2}, \pm {\pi \over 
2})$. The MF $\bar B$ 
 turns the spinless fermion 
$H$ into two species of 2--component Dirac fermions  $\psi^{(r)},r=1,2$, 
each of them being supported on one N\'eel sublattice.
The continuum action for these fermions is given by:$^{\cite{mar1}}$
\begin{equation}
\int dx_0 d^2 x \sum_r \bar\psi^{(r)} 
[\gamma^0(\partial_0 - e_r A_0 - \delta) + t (\parz - e_r \A)]\psi^{(r)}, \label{holon}
\end{equation}
where $ \A = \gamma_\mu A_\mu, \parz = \gamma_\mu  \partial_\mu,
\gamma_0 =\sigma_z, \gamma_\mu = (\sigma_y, \sigma_x)$,
  the charges $e_r= \pm 1$, depending on sublattice.
After  integrating out the gapful Dirac modes, we end up with
FL-like system of holons  with Fermi energy $\epsilon_F \sim t\delta$, 
interacting through
gauge field $A$. 

As shown in \cite{il},  for the  gauge field model 
the  in--plane resistivity is  approximately given by
\begin{equation}
R = {\lim\limits_{\omega \rightarrow 0}} \omega [({\rm Im} 
\Pi_s^\bot ( 
\omega))^{-1} + ({\rm Im} \Pi_h^\bot (\omega))^{-1}], \label{resis}
\end{equation}
where $\Pi^\bot_s$ and $\Pi^\bot_h$ denote the transverse 
polarization bubbles (at $\vec q=0$) due to the $h/s$ currents of holons and spinons, 
renormalized  by    gauge   fluctuations. The $A$ 
propagator  for small $|\vec q|, \omega, \omega/|\vec q|$ 
 in the Coulomb gauge is given by:$^{\cite{ln,fro2}}$
$$
\langle A^\bot_\mu A^\bot_\nu \rangle (q, \omega) \sim (i \omega \lambda_h 
(\vec q) + \chi |\vec q|^2)^{-1}, 
$$
\begin{equation}
\quad
\langle A_0 A_0 \rangle (q, \omega) \sim (\nu_h + \omega_p)^{-1},
\label{gaugepro}
\end{equation}
where $\lambda_h \sim \kappa / |\vec q|, \kappa \sim 0 (\delta)$ is the Landau 
damping due to a finite FS for holons, $\chi = \chi_h + 
\chi_s, \chi_h \sim m_h^{-1} \sim O (\delta^{-1}), \chi_s = v_s m_s^{-1} \sim 
O ((-\delta \ln \delta)^{-{1\over 2}})$ is the diamagnetic susceptibility,
$\nu_h$   is the holon density at the FS and $\omega_p$ 
is the  plasmon gap.

An estimate of the holon contribution to resistivity can be derived as in 
\cite{ln}, 
\begin{equation}
R_h 
\sim 
(\epsilon_F  \tau_{imp})^{-1} + ({T\over \chi})^{4/3}/\epsilon_F,
\label{holonr}
\end{equation}
where $\tau_{imp}$ is the  transport relaxation time due to impurities.

To estimate the spinon contribution, we derive the large 
scale behavior of the spinon current 
$j^{\mu}=z^*D_A^{\mu} z$   correlation function, where 
$D_A^{\mu}=\partial_{\mu}-A_{\mu}$,
by eikonal approximation,$^{\cite{fradk}}$ strictly preserving gauge invariance. 
We use spinon Green functions at zero 
temperature, as partially justified by the spinon gap, but we retain the 
temperature dependence of gauge fluctuations.

We apply the Fradkin representation$^{\cite{fradk}}$  to the spinon propagator
$\langle z(x) z^*(y) \rangle =G(x,y|A)$. 
It  can be 
  derived using a first--quantized path integral form
of the propagator, with  metric $(-++)$,  replacing 
integration over trajectories $q_\mu (t)$ by integration over 3--velocities
$\phi_\mu = \dot q_\mu (t), \mu= 0,1,2$. Rescaling $x_0$ to $v_s x_0$ 
one obtains:
$$
G (x,y|A) = i \int^\infty_0 ds {\rm e}^{-is m^2} [e^{is (\partial_\mu - 
A_\mu)^2} ] (x,y) 
$$
$$
\sim i \int^\infty_0 ds {\rm e}^{-ism^2} \int {\cal D} \phi^\mu (t) {\rm e}^{{i\over 4}
\int^s_0 \phi^2_\mu (t) dt}
$$
\begin{equation}
\cdot e^{i\int^s_0 \tilde{A}_\mu(t) \phi^\mu (t) dt}
\int d^3 p e^{ip_\mu (x^\mu - y^\mu - \int^s_0 \phi^\mu (t) dt)}. \label{frad}
\end{equation}
Using an identity  (Eq. (41) in the second paper of \cite{fradk}),
the integral $\int^s_0 \tilde{A}_\mu \phi^\mu (t) dt$  with $ \tilde{A}_\mu =
A_\mu(x+\int_0^t \phi(t^\prime) dt^\prime)$ can be decomposed into a sum of 
an integral along a straight line (denoted by $\int_x^y A$) and a gauge invariant part
depending on the field strength $F_{\mu\nu}$.
Thus
$
G(x,y|A)=\exp\{-i \int_x^y A\} G(x,y|F).
$
The spinon current density correlation $\langle j^{\mu} (x) j^{\mu}(y) \rangle$, 
$\mu=1,2$, is approximately given by $\langle 
D_A^{\mu}(x)G (x,y|A)
D_A^{\mu}(y)G (x,y|-A) \rangle$, where $\langle \cdot \rangle$ denotes  
average w.r.t. $A$.
The gauge--dependent terms
of  the two spinon propagators  exactly cancel each other, yielding a strictly
gauge--invariant result, at large scale given approximately by 
$
\langle 
{\partial \over \partial x_{\mu}} G(x,y|F) {\partial \over \partial 
y_{\mu}} G(x,y|-F) \rangle.
$ 
The $A^\bot$-- average involves contributions 
weighted by ``magnetic field" correlations $\langle F_{\mu\nu} (z)
F_{\rho\sigma} (w) \rangle,\mu,\nu,\rho,\sigma= 1,2$, approximately 
evaluated for $|z^0- w^0| \ll T^{-1}$ as in \cite{ln,fro2}, obtaining 
$
(\delta_{\mu\rho} \delta_{\nu\sigma} -  \delta_{\mu\sigma} 
\delta_{\nu\rho})
{4 T\over \chi} e^{-|\vec z -\vec w|^2 q^2_0} q^2_0,
$
where $q_0= ({\kappa\over \chi\beta})^{1\over 3}$ is a momentum cutoff related 
to the anomalous skin effect due to the Reizer singularity in the $A^\bot$ 
propagator.$^{\cite{ln}}$ The $A_0$ average involves contributions weighted by 
``electric--field" correlations
$
\langle F_{0\mu} (z) F_{0\nu} (w) \rangle, \mu,\nu =1,2.
$
Since they vanish in the limit $q, \omega \sim 0$ 
(see (\ref{gaugepro})), their contributions will be neglected.

Integrals in (\ref{frad}) can be approximately calculated for relatively 
low temperatures  ($ T < \chi m_s^2$) and the current-current correlation becomes:
$$
\langle j^{\mu} (x) j^{\mu} (0) \rangle\sim \bigl[ {\partial \over 
\partial x_{\mu}} {\rm e}^{-i(x^2_0 -|\vec x|^2)^{1\over 2} 
(m^2 -  {T\over \chi} f (1/2 |\vec x| q_0))^{1\over 2} }
$$
\begin{equation}
\cdot {\rm e}^{-{T q^2_0\over 4 \chi} { g(1/2 |\vec x| q_0) \over m^2 } (x_0^2-
|\vec x|^2)}(x_0^2 - |\vec x|^2)^{-1/2} \bigr]^2 
\label{jj1}
\end{equation}
where, for a real argument, $f$ is monotonically 
increasing, vanishing at zero argument and $g$ 
is monotomically decreasing, vanishing at large arguments.
Their explict expressions are  lengthy and will be given elsewhere.$^{\cite{marc2}}$
In deriving the 
spinon current correlation at $\vec q = 0$ we 
evaluate the $\vec x$--integration by saddle point. For
$x_0 \gg q_0^{-1}$ the integral 
is dominated by a complex saddle point at $|\vec x|=2 q_0^{-1} 
\alpha(x_0)$, with finite $\alpha (x_0)$ (in the first quadrant),
having a weak dependence on $x_0$.
To justify the saddle point approximation we need to assume 
 $T > \chi m_s q_0$.  It turns out that in the physical range of
parameters considered in the paper, this and the above  conditions are both satisfied
for   temperatures between   tens   and a few hundred degrees.

\begin{figure}
\epsfxsize=3.3 in \centerline{\epsffile{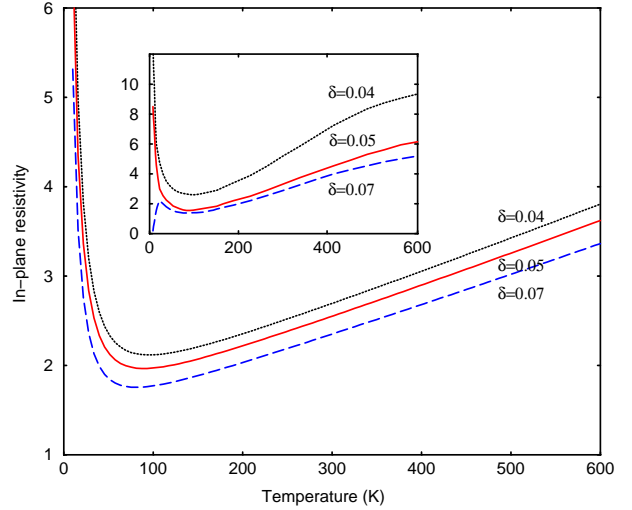}}
\caption[]{The calculated temperature dependence of in-plane resistivity
(sum of (\ref{holonr}) and (\ref{spinonr})) for various
 dopings $\delta$ 
in comparision with the corresponding experimental data (inset) 
on La$_{2-\delta }$Sr$_{\delta}$CuO$_4$ in units of 
$m\Omega cm$,  taken from \cite{takagi}.}  
\label{fig1}
\end{figure}

\begin{figure}
\epsfxsize=3.3 in \centerline{\epsffile{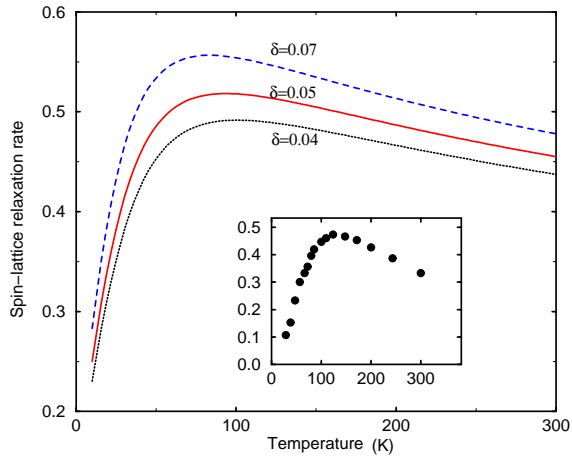}}
\caption[delfsa]{The temperature dependence of calculated spin-lattice relaxation rate
$(T_1T)^{-1}$ given by (\ref{t1r}).  Inset: $^{63}(T_1T)^{-1}$ in the CuO$_2$ planes of
YBa$_2$Cu$_3$O$_{6.52}$  single crystals in units of $s^{-1}K^{-1}$,  taken from \cite{ber}. }
\label{fig2}
\end{figure}

Let us define 
$
\Pi^+(\omega)=\int^{\infty}_0 dx_0 \langle j^{\mu} j^{\mu} 
\rangle^{\star}(\vec q =0, x_0) e^{i x_0 \omega}.
$
Using the Lehmann representation we find 
$
\lim_{\omega \rightarrow 0} {\rm Im} \Pi^{\bot}(\omega) \omega^{-1}=-2{\partial 
\over \partial \omega} {\rm Re} \Pi^+ (0).
$
Taking note that the main contribution comes from small 
$x^0$, introducing lower cut-off and performing scale renormalization, we obtain for 
the  $\omega \to 0 $ limit:
 $
\partial
 {\rm Re} \Pi^+ (\omega)/ \partial \omega
$
$$ \sim
{\rm Re}  \left[ (\alpha(0))^3
(q_0)^{3/2} {\cal Z}^{1/4}
( i f^{\prime\prime} )^{-1/2}
({T \over  \chi})^{-1/2}
(\omega-{\cal Z}^{1/2})^{-1}\right],
$$
where 
\begin{equation}
{\cal Z} = |{\cal Z}| {\rm e}^{ -i \theta} \equiv m^2 - {T\over \chi} f(\alpha(0)),\;\; 
 f^{\prime\prime} \equiv
 f^{\prime\prime} (\alpha(0)). \label{notat}
\end{equation}
(renormalization eliminates the contribution of the $g$ function,
being subleading).
We find that at large $x_0$  $\arg \alpha(0)={\pi \over  4}$  
and $\arg f''(\alpha(0))=0$. We extrapolate  $\alpha (x_0)$ to  $\alpha
(0)$, 
keeping these features. This way  we 
recover the correct behavior, $R \rightarrow \infty,$ as $T \rightarrow 0$. 
As $x_0 \rightarrow 0$,
the saddle point extrapolates to $x_s \sim  q_0^{-1} e^{i {\pi \over 4}}$,
and we find  the ``spinon contribution'' to resistivity:
\begin{equation}
R_s = 2^{-4}  \bigl( { |f^{\prime\prime} |\over \kappa} \bigr)
 ^{1\over 2} |\alpha(0)|^{-3} {|{\cal Z}|^{1/4}
\over \sin(\theta / 4) }.
\label{spinonr}
\end{equation}
In Fig. \ref{fig1}  our calculated resisitivity (sum of (\ref{holonr}) and (\ref{spinonr})),
is plotted as a function of temperature for various dopings in comparison
with experimental data taken on LSCO$^{\cite{takagi}}$ (inset). We have taken
$t/J =3, J= 0.1$ eV. Apart from the resistivity scale, there are no other
adjustable parameters (similarly for Fig. \ref{fig2}). We find a resistivity minimum
 below 100K in
very good agreement with experiment. We see from (\ref{notat}) that the imaginary part of 
$\cal Z$ is proportional to temperature $T$. At low temperatures the
spin gap
effect ($\sim m_s$) dominates, $\theta \rightarrow 0$, so the system shows
an  insulating behavior. (The functional dependence $ R \sim 1/T$,  different
from the ``standard'' exponential law due  to  spin gap, is a  prediction
of our theory.)   On the contrary, at higher temperatures the imaginary and real
parts of $\cal Z$ become comparable, so the resistivity
 grows with temperature
due to  gauge  fluctuations via $|\cal Z|$. Moreover, the minimum 
shifts to higher temperatures, as the doping decreases, also in agreement with
experiment ( our theoretical prediction $ m_s^2 \sim - \delta \ln \delta$, rather than $ \sim \delta$
is responsible for this shift). We have also compared the calculated conductivity
in the semi-log scale  with data taken on
a very good single crystal of La$_{1.96}$Sr$_{0.04}$CuO$_4$ (inset).$^{\cite{kas,pre}}$ We 
find a symmetric shape of curve around the maximum, and a reflection point as well as
a linear piece on the low temperature side  in both theory and experiment. 
So far we have not included the external
magnetic field. We believe the experimentally  observed crossover from
metallic to insulating behavior in strong magnetic fields when superconductivity
is suppressed,$^{\cite{ando}}$ can be understood in a similar way, and this issue
will be addressed in our future communication.$^{\cite{marc2}}$

Now  turn to the spin--lattice relaxation rate $T_1^{-1}$ which can be expressed 
approximately  as:$^{\cite{ber}}$
$$
(T_1 T)^{-1} \sim {\lim\limits_{\omega\rightarrow 0}} \int d^2 q {\cal F}(\vec q) 
{{\rm Im} \chi_s (\vec q, \omega) \over \omega},
$$
where $\chi_s$ is the spin susceptibility and ${\cal F} (\vec q)$ is the form
 factor.
To evaluate $\chi_s$ we use the representation for spin deduced at large 
scales
$
\vec S_x \sim e^{i\pi|x|} z^* \vec\sigma z (x) (1 - \rho_h (x)),
$
where $\rho_h$ is the holon density, to be replaced
by its
\noindent  average $\delta$. 
Around the 
AF wave vector 
${\cal\vec Q} _{AF} = (\pi,\pi)$ we find that
$
\langle \vec S(x) \cdot \vec S(0)\rangle \sim
(1-\delta)^2 e^{i \pi |\vec x|} \langle G(x,0|F) G(x,0|-F) \rangle
$
which can be calculated as before.
Define
$
\chi^+(\vec q,\omega)= \int_0^\infty dx_0 \langle \vec S \cdot \vec S
\rangle ^\star (\vec q, x_0) e^{i x_0 \omega},
$
using the Lehman representation and taking into account that ${\cal F} (q)$
is even in $\vec q$ one obtains
$
(T_1 T)^{-1} =  -2 \int d^2 q {\cal F} (\vec q) { \partial {\rm Re }\chi^+ ( {\vec q}, 
\omega =0) \over {\partial \omega}}.
$
Since ${\cal F }(\vec q) $ is peaked around ${\cal\vec Q} _{AF}$ for Cu,
integrating over $q$ in a small region around that point,
and introducing a cutoff in the real space $\Lambda \sim \pi / |x_s|$,
we find  
\begin{equation}
(T_1 T)^{-1} \sim (1-\delta)^2 \sqrt{\delta} |{\cal Z}|^{- {1 \over 4}} (a \cos ({ \theta \over 4})
+ b \sin ({\theta \over 4})),\label{t1r}
\end{equation}
 $ a= {\rm Re}\int_\Lambda d^2 y J_0 (2|\vec y|\alpha (0)),
b= - {\rm Im} \int_\Lambda d^2 y J_0 (2|\vec y|\alpha (0))$, 
and $ J_0$ is the zero-order Bessel function.
In Fig. \ref{fig2} we plot our calculated spin-lattice relaxation rate $(T_1T)^{-1}$ for $^{63}$Cu
as a function of temperature for various dopings in comparison with experimental data 
taken on underdoped samples of YBCO.$^{\cite{ber}}$ We observe a maximum near the
crossover temperature for conductivity, although the shape around  maximum is not
symmetric anymore, due to the presence of the cosine term.

\begin{figure}
\epsfxsize=2.5 in \centerline{\epsffile{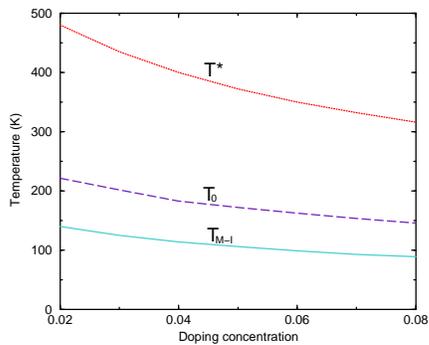}}
\caption[delfsa]{The calculated metal-insulator crossover temperature $T_{M-I}$,
the reflection point  $T_0$ in NMR $(T_1 T)^{-1}$, and the reflection point $T^*$ of $R$
as functions of doping. }
\label{fig3}
\end{figure}

To summarize, we have shown using the $U(1)\times SU(2)$ gauge field theory
that the metal-insulator behavior crossover and peculiar behavior of NMR
relaxation in underdoped cuprates might be due to the interplay of the spin gap
(derived in our approach) effect and the gauge field fluctuations. More precisely,
the crossover is taking place when the real and imaginary parts of ${\cal Z}$
(Eq. (\ref{notat})) become  comparable. In Fig. \ref{fig3} we have plotted 
 three different crossover temperatures related to the spin gap effects,
namely the metal-insulator crossover $T_{M-I}$ (minimum
of the in-plane resistivity), the spin gap crossover temperatures, detected
by NMR $T_0$ and by resistivity $T^*$, identified with their respective  reflection points,
 in the low-doping region ( $\delta \sim 0.02 - 0.08$). The last two temperatures.
roughly speaking, limit from the above the region  of  significant spin gap effects and validity of
 our approximation. These  crossover
temperatures  are different manifestations of the same energy scale.
The fact that their relative order $T^*  >T_0 > T_{M-I}$,
as well as their order of magnitude agrees with experiments, provides
further support for  our approach.


 The work of P.M. was partially supported by
TMR Programme ERBFMRX-CT96-0045.

\end{document}